# Results on charged kaon and hyperon decays from NA48

Monica Pepe (for the NA48 collaboration)
*INFN, Sezione di Perugia, via A. Pascoli, 06123 Perugia, ITALY*

Recent results from the NA48/1 and NA48/2 experiments at the CERN SPS are presented. NA48/2 carried out data taking in 2003 and 2004 collecting charged kaon decays: branching ratios and form factors have been measured for the rare $K^\pm \to \pi^\pm e^+ e^-$, $K^\pm \to \pi^\pm \gamma \gamma$ and $K^\pm \to \pi^\pm e^+ e^- \gamma$ decays. The NA48/1 experiment has taken data in 2002 using only the $K_S$ beam at an increased intensity measuring neutral hyperon radiative decays. Using this data, which exceeds present statistics by about one order of magnitude, a new precise measurement of the $\Xi^0$ decay asymmetries has been obtained.

## 1. INTRODUCTION

The main goal of NA48/2 was the search for direct CP violation in $K^\pm$ decays into three pions. However, given the high statistics achieved, many other physics topics were also covered. When the NA48/1 Collaboration undertook investigations with a high-intensity $K_S$ beam in 2002, trigger strategies for identifying radiative hyperon decays were also included.

The paper is organized as follows: Section 2 describes the NA48/2 experiment and presents recent results on rare charged kaon decays, while Section 3 is devoted to NA48/1 results on radiative hyperon decays.

## 2. THE NA48/2 EXPERIMENT

Simultaneous $K^+$ and $K^-$ beams were produced by 400 GeV/c primary protons from the CERN SPS, impinging on a Be target. Charged particles with momentum $(60 \pm 3)$ GeV/c were selected by an achromatic system of four dipole magnets ('achromat'), which splits the two beams in the vertical plane and then recombines them on a common axis.

The main components of the NA48 detector are the magnetic spectrometer, consisting of four drift chambers and a central dipole magnet for charged particle reconstruction, and the liquid krypton electromagnetic calorimeter (LKr), an almost homogeneous ionization chamber with an active volume of 10 m$^3$, used to measure electromagnetic showers. In addition, the detector is comprised of a hodoscope for precise track time determination, a hadronic calorimeter, a muon detector and a large angle photon veto system. A more detailed description of the detector can be found in [1].

### 2.1. $K^\pm \to \pi^\pm e^+ e^-$ decay

Radiative non leptonic kaon decays represent a source of information on the structure of the weak interactions at low energies, providing crucial tests of the Chiral Perturbation Theory (ChPT). The FCNC process $K^\pm \to \pi^\pm e^+ e^-$, induced at one-loop level in the Standard Model and highly suppressed by the GIM mechanism, has been described by the ChPT [2] and several models predicting the form factor characterizing the dilepton invariant mass spectrum and the decay rate have been proposed [3,4]. This decay was first studied at CERN [5], followed by BNL E777 [6] and E865 [7] measurements. The $K^\pm \to \pi^\pm e^+ e^-$ decay rate is measured with respect to the abundant $K^\pm \to \pi^\pm \pi^0_D$ normalization channel: having both final states identical sets of charged particles, electron and pion identification efficiencies, potentially representing a significant source of systematic uncertainties, cancel at first order. The reconstructed $\pi^\pm e^+ e^-$ invariant mass spectrum is presented in Figure 1 (left).

In total 7146 $K^\pm \to \pi^\pm e^+ e^-$ candidates are found in the signal region with residual background contamination of 0.6%, mostly resulting from particle misidentification. A preliminary model independent measurement of the branching ratio (BR) for $z = M^2_{ee}/M_K > 0.08$ gave BR = $(2.26 \pm 0.03_{stat} \pm 0.03_{syst} \pm 0.06_{ext}) \cdot 10^{-7}$. Model dependent fits to the z-spectrum have been performed (Figure 1 right), obtaining the corresponding form factors and BR: the data sample is insufficient to distinguish between the different models. The preliminary average BR in the full kinematic range is BR = $(3.08 \pm 0.04_{stat} \pm 0.04_{syst} \pm 0.08_{ext} \pm 0.07_{model}) \cdot 10^{-7}$.

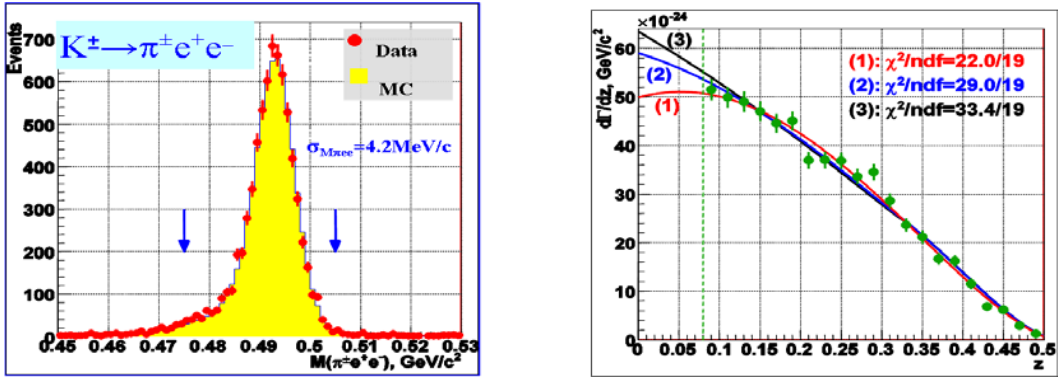

Figure 1: Reconstructed spectrum of $\pi^\pm e^+ e^-$ invariant mass (Left) and the computed $d\Gamma_{ee}/dz$ (background and efficiency corrected, Right) compared to fit results according to the considered models (1=linear, 2=[3], 3=[4]).

## 2.2. $K^\pm \to \pi^\pm \gamma \gamma$ and $K^\pm \to \pi^\pm e^+ e^- \gamma$ decays

The $K^\pm \to \pi^\pm \gamma\gamma$ and $K^\pm \to \pi^\pm e^+ e^- \gamma$ decays similarly arise at one-loop level in the ChPT. The decay rates and spectra have been computed at leading and next-to-leading orders [8,9] and strongly depend on a single theoretically unknown parameter $\hat{c}$. The experimental knowledge of these processes is rather poor: before the NA48/2 experiment, only a single observation of 31 $K^\pm \to \pi^\pm \gamma\gamma$ events was made [10], while the $K^\pm \to \pi^\pm e^+ e^- \gamma$ decay was not observed at all.

The $K^\pm \to \pi^\pm \gamma\gamma$ rate is measured relatively to the $K^\pm \to \pi^\pm \pi^0$ normalization channel. About 40% of the total NA48/2 data sample have been analyzed, finding 1164 signal candidates with 3.3% background. The reconstructed spectrum of $\gamma\gamma$ invariant mass in the accessible kinematic region M > 0.2 GeV/$c^2$ is presented in Figure 2 (left), along with a MC expectation assuming ChPT O($p^6$) distribution [8] with a realistic parameter $\hat{c} = 2$. ChPT predicts an enhancement of the decay rate (cusp-like behavior) at the $\pi\pi$ mass threshold $m_{\gamma\gamma} = 280$ MeV/$c^2$, independently of the value of the $\hat{c}$ parameter. The observed spectrum provides the first clean experimental evidence for this phenomenon.

The model dependent BR has been measured under the above assumptions: the preliminary result is BR = $(1.07 \pm 0.04_{stat} \pm 0.08_{syst}) \cdot 10^{-6}$. A model independent BR measurement is in preparation, together with the extraction of $\hat{c}$ from a combined fit to the mass spectrum and branching ratio.

The $K^\pm \to \pi^\pm e^+ e^- \gamma$ decay is similar to $K^\pm \to \pi^\pm \gamma\gamma$ with one photon internally converting into a pair of electrons. NA48/2 has reported the first observation [11] of the decay $K^\pm \to \pi^\pm e^+ e^- \gamma$ using the full 2003 and 2004 data sample: 120 signal candidates with 7.3±1.7 estimated background events have been selected in the accessible region with $M_{ee} > 0.26$ GeV/$c^2$ invariant mass. The candidates are shown in Figure 2 (right). Using $K^\pm \to \pi^\pm \pi^0_D$ as normalization channel, the BR has been determined in a model independent way: BR = $(1.19 \pm 0.12_{stat} \pm 0.04_{syst}) \cdot 10^{-8}$ for $M_{ee} > 0.26$ GeV/$c^2$.

The $\hat{c}$ parameter has also been measured assuming the validity of O($p^6$) [9] and found to be $\hat{c} = 0.90 \pm 0.45$.

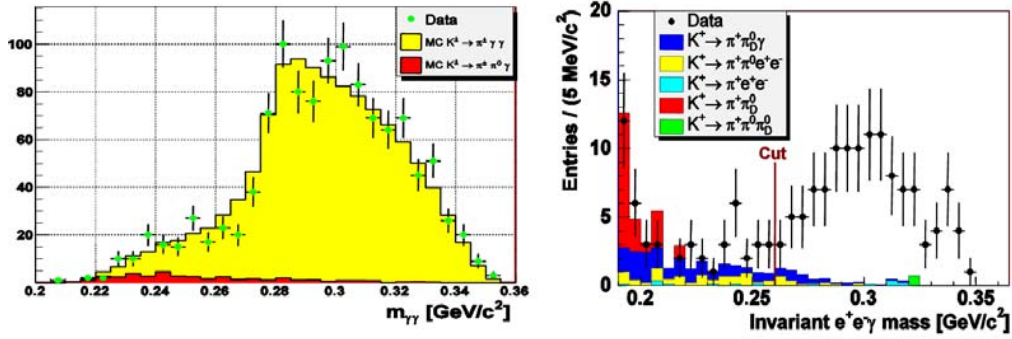

Figure 2: Reconstructed spectrum (dots) of $\gamma\gamma$ invariant mass for $K^\pm \to \pi^\pm \gamma\gamma$ decays (Left) and of $\gamma e^+e^-$ invariant mass for $K^\pm \to \pi^\pm e^+e^-\gamma$ decays (Right) compared to MC expectations (filled areas).

### 3. THE NA48/1 EXPERIMENT

The NA48 beam line was originally designed to produce and transport both $K_L$ and $K_S$ beams simultaneously. In the 2002 run, using only the $K_S$ beam at an increased intensity, NA48/1 was able to measure rare $K_S$ and neutral hyperon decays with a total flux of more than 3 billions $\Xi^0$ decays. The detector was the same as described in Section 2.

### 3.1. $\Xi^0 \to \Lambda\gamma$ and $\Xi^0 \to \Sigma^0\gamma$ decays

The precise nature of weak radiative hyperon decays such as $\Xi^0 \to \Lambda\gamma$ and $\Xi^0 \to \Sigma^0\gamma$ is still barely understood. Several theoretical models [12,13] exist which give different predictions, and an excellent experimental parameter to distinguish between models is the decay asymmetry $\alpha$ of these decays, defined as $dN/d\cos(\theta) = N_0 (1 + \alpha \cdot \cos(\theta))$, where $\theta$ is the angle between the direction of the daughter baryon flight and the polarization of the $\Xi^0$ in its rest frame. As an example, the decay asymmetry for $\Xi^0 \to \Lambda\gamma$ can be measured by looking at the angle between the incoming $\Xi^0$ and the outgoing proton from the subsequent $\Lambda \to p\pi^-$ decay in the $\Lambda$ rest frame, making the measurement independent of the unknown initial $\Xi^0$ polarization.

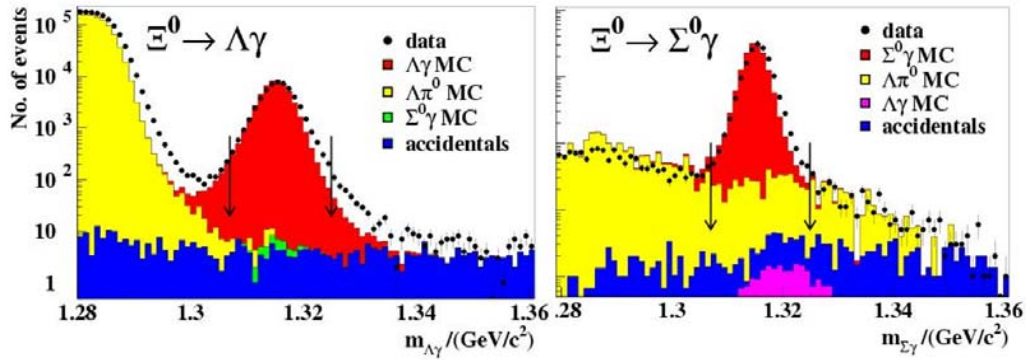

Figure 3: $\Xi^0 \to \Lambda\gamma$ (Left) and $\Xi^0 \to \Sigma^0\gamma$ (Right) candidates (dots) compared to MC expectations (filled areas) for signal and backgrounds.

The NA48/1 experiment has selected 48314 $\Xi^0 \to \Lambda\gamma$ and 13068 $\Xi^0 \to \Sigma^0\gamma$ candidates (Figure 3), with 0.8% and 3% background contribution respectively. In the case of $\Xi^0 \to \Sigma^0\gamma$, with the subsequent decay $\Sigma^0 \to \Lambda\gamma$, the product $cos(\theta_{\Xi \to \Sigma\gamma}) \cdot cos(\theta_{\Sigma \to \Lambda\gamma})$ has to be used for the fit to the decay asymmetry. Both fits show the expected linear behaviour on the angular parameters; after correcting for the well known asymmetry of $\Lambda \to p\pi^-$ the following preliminary results are obtained: $\alpha_{\Xi\Lambda\gamma} = -0.684 \pm 0.020_{stat} \pm 0.061_{syst}$ and $\alpha_{\Xi\Sigma\gamma} = -0.682 \pm 0.031_{stat} \pm 0.065_{syst}$ in agreement with previous measurements [14,15] but with higher precision. The large negative value of the $\Xi^0 \to \Lambda\gamma$ decay asymmetry, difficult to accommodate for quark and vector meson dominance models, is confirmed.

### 3.2. $\Xi^0 \to \Lambda e^+ e^-$ decay

The weak radiative $\Xi^0 \to \Lambda e^+ e^-$ decay has been detected for the first time. Assuming an inner bremsstrahlung-like mechanism producing the $e^+e^-$ pairs, a naïve estimation of the expected rate for this process is obtained considering a (virtual) photon internal conversion (Dalitz decay) or using QED predictions as for the rate of $\Sigma^0 \to \Lambda e^+ e^-$ [16].

In the signal region 412 candidates have been found, with an estimated background of 15±5 events. The branching ratio BR = $(7.6 \pm 0.4_{stat} \pm 0.4_{syst} \pm 0.2_{norm}) \cdot 10^{-6}$ has been determined [17], consistent with an internal bremsstrahlung process. The decay asymmetry parameter, measured from the angular distribution of the proton relative to the $\Xi^0$ line of flight in the $\Lambda$ rest frame, is $\alpha_{\Xi\Lambda ee} = -0.8 \pm 0.2$ consistent with that of $\Xi^0 \to \Lambda\gamma$.